# Two-Color-Laser-Driven Direct Electron Acceleration in Infinite Vacuum

**Liang Jie Wong\* and Franz X. Kärtner**

*Department of Electrical Engineering and Computer Science and Research Laboratory of Electronics, Massachusetts Institute of Technology, 77 Massachusetts Avenue, Cambridge, MA, 02139, USA*
*\*Corresponding author: ljwong@mit.edu*

We propose a direct electron acceleration scheme that uses a two-color pulsed radially-polarized laser beam. The two-color scheme achieves electron acceleration exceeding 90% of the theoretical energy gain limit, over twice of what is possible with a one-color pulsed beam of equal total energy and pulse duration. The scheme succeeds by exploiting the Gouy phase shift to cause an acceleration-favoring interference of fields only as the electron enters its effectively final accelerating cycle. Optimization conditions and power scaling characteristics are discussed.

Laser-driven electron acceleration has been receiving increasing attention in recent years due to its potential to realize tabletop GeV electron accelerators, which have a broad range of applications in medicine and science. Although plasma-based acceleration schemes [1] have had much experimental success, the possibility of accelerating electrons in vacuum [2,3] remains of great interest since the absence of plasma would preclude problems associated with the inherent instability of laser-plasma interactions. Among the myriad vacuum acceleration schemes proposed is the direct acceleration of electrons in infinite vacuum by a pulsed radially-polarized laser beam. In this scheme, electrons near the beam axis are accelerated primarily by the laser pulse's longitudinal electric field component. Such a scheme is appealing due to the low radiative losses of direct acceleration [4], the absence of limits on laser field intensity and electron confinement to the vicinity of the beam axis, which favors the production of mono-energetic and well-collimated electron beams [5-7]. Simulations reveal that electron acceleration to MeV energies is already possible with peak laser powers as low as a few TW, and that much more dramatic energy gains may be achieved with initially-moving ("pre-accelerated") electrons [8].

Previous simulations have shown that a one-color pulsed radially-polarized laser beam can accelerate an initially-stationary electron only up to 40% of the theoretical energy gain limit [8,9]. In this Letter, we show that a two-color pulsed beam can accelerate an electron by over 90% of the one-color beam's theoretical gain limit, for a given total energy and pulse duration. The scheme exploits how the Gouy phase shift will vary the interference pattern of the on-axis electric field with position along the beam axis. For most cases well above the threshold power for electron acceleration, maximum acceleration is obtained with an acceleration-favoring interference of fields only as the electron enters its effectively final accelerating cycle.

The two-color pulsed beam is the sum of two co-propagating pulsed radially-polarized laser beams, with central angular frequencies $\omega$ and $2\omega$, of equal pulse duration, peak power and Rayleigh range. The electron begins at rest on the beam axis in field-free vacuum (the pulse begins infinitely far away) and ends moving in field-free vacuum after the pulse has completely overtaken it (the setup is identical to that in [8], with the one-color beam replaced by a two-color beam). On the beam axis all transverse fields vanish, leaving the longitudinal electric field $E_z$, which is obtained by summing the $E_z$ components of two one-color beams:

$$E_z = \sqrt{4\eta_0 P/\pi} \Big/ \big[z_0\big(1+(z/z_0)^2\big)\big] \mathrm{sech}\big((\xi+kz_i)/\xi_0\big) \cdot$$
$$\big\{\sin\big((\xi+\psi_a)+\psi_g+\psi_b\big) + \sin\big(2(\xi+\psi_a)+\psi_g+\psi_b\big)\big\} \quad (1)$$

where $\xi \equiv \omega t - kz$; $z_0 \equiv \pi w_0^2/\lambda$ is the Rayleigh range; $k \equiv 2\pi/\lambda = \omega/c$; $w_0$ is the waist radius of the fundamental harmonic beam; $\psi_g \equiv 2\tan^{-1}(z/z_0)$ is the Gouy phase shift; $\eta_0 \cong 120\pi\,\Omega$ is the vacuum wave impedance; $z_i$ is the pulse's initial position (effectively $-\infty$); $c$ is the speed of light in vacuum; $\psi_a$ and $\psi_b$ are phase constants; $\xi_0 \equiv \omega\tau/\mathrm{sech}^{-1}(\exp(-1))$, where $\tau$ is the pulse duration; $P/2$ is the peak power of each pulse; $z(0)$ is the initial electron position.

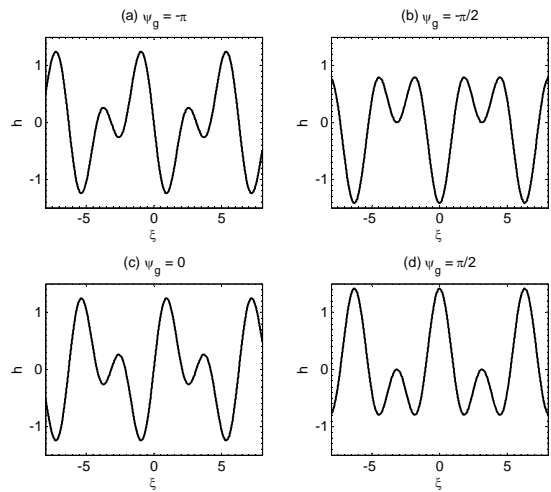

Fig. 1. Plots of $h \equiv \sqrt{0.5}\big[\sin(\xi+\psi_g) + \sin(2\xi+\psi_g)\big]$ for various $\psi_g$.

Our results should closely approximate those for the more general case of a slightly off-axis, non-relativistic electron, due to the electron confinement property of the transverse fields [5, 6] and the fact that the laser pulse and phase move at or beyond the speed of light.

Consider $h \equiv \sqrt{0.5[\sin(\xi+\psi_b)+\sin(2\xi+\psi_g)]}$, to which Eq. (1) is proportional except for a translation in $\xi$ and $\psi_g$, plotted in Fig. 1. The phase $\psi_b$ in Eq. (1) controls the field pattern produced by interference at each position along the beam axis. For instance, setting $\psi_b = \pi$ would cause the field pattern to evolve, due to the Gouy phase shift, in the order (c)-(d)-(a)-(b)-(c) as the laser pulse propagates from $z = -\infty$ to $-z_0$, 0, $z_0$ and $\infty$ respectively. We also note that of all possible patterns, the one in Fig. 1(b) seems to favor electron acceleration most, since its ratio of most negative to most positive value is largest in magnitude. The position where the Fig. 1(b) wave pattern occurs is given by $z_b = -z_0 \tan(\pi/4 + \psi_b/2)$.

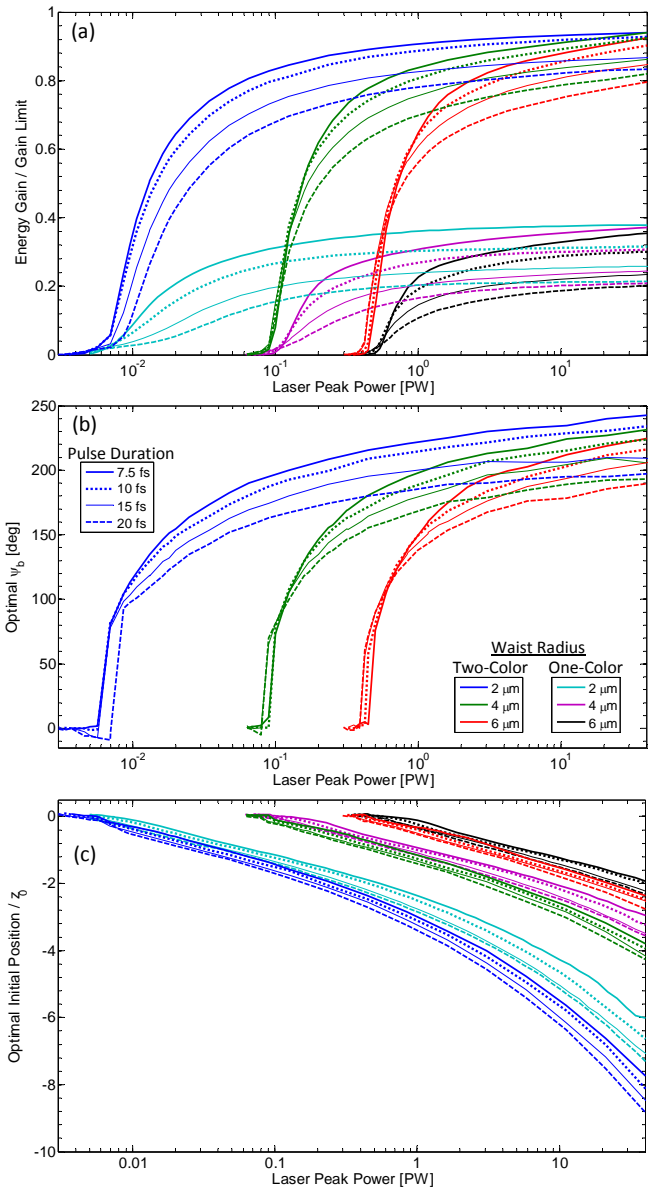

Fig. 2. Plots of (a) maximum normalized energy gain (b) corresponding optimal $\psi_b$ (for two-color case) and (c) corresponding optimal normalized $z(0)$ vs. peak power.

We numerically solve the Newton-Lorentz equations of motion [8] using the Adams-Bashforth-Moulton method (*ode113* of *Matlab*). Although we set $\lambda = 0.8$ μm here, our results are readily scalable to any $\lambda$ since the electrodynamic equations are independent of $\lambda$ under the normalizations $T \equiv \omega t$, $\varsigma \equiv z/z_0$ (with $\varsigma_i \equiv z_i/z_0$) and $\kappa \equiv kz_0 = 2(\pi w_0/\lambda)^2$ (as was explicitly shown for the one-color beam in [8]). We sweep over $P$-$\tau$-$w_0$ space and optimize over $\psi_a$-$\psi_b$-$z(0)$ space for electron energy gain normalized by the one-color theoretical energy gain limit $\Delta E_{\lim} = e(8\eta_0 P/\pi)^{1/2} \cong (P/[PW])^{1/2} [GeV]$ ($P/[PW]$ denoting $P$ in units of petawatts) [8,9]. The gain limit is computed by considering an electron that (unrealistically) remains at the pulse peak and in one accelerating cycle from the focus to infinity. As Fig. 2(a) shows, the two-color beam with peak power $P/2$ in each beam, and therefore the same total power as the one-color beam ($\lambda = 0.8$, as in [8]) with peak power $P$, can accelerate an electron by more than 90% of the one-color beam's theoretical gain limit, whereas the one-color beam can manage less than 40% in the parameter space studied. Fig. 2(b) shows that well above the threshold power, the optimal $\psi_b$ lies between $\pi/2$ and $3\pi/2$, i.e. the Fig. 1(b) wave pattern occurs between $z = 0$ and $z = \infty$, with a tendency to be around $\pi$ (the Fig. 1(b) wave pattern occurs around $z = z_0$). This accords with physical intuition because a) due to the Gouy phase shift the electron can enter its effectively final accelerating cycle only after $z = 0$ and b) when determining the best position for the Fig. 1(b) pattern, one must strike a compromise between the Lorentzian decay (due to beam divergence) in Eq. (1) and the fact that the acceleration-favoring Fig. 1(b) pattern will be maintained over a greater distance the further from the focus it occurs, due to the smaller rate of change with distance of the Gouy phase shift. Fig. 2(c) shows that the optimal initial position of the initially-stationary electron for the two-color beam tends to be slightly more negative than that for the one-color beam with the same peak power $P$, pulse duration $\tau$ and waist radius $w_0$.

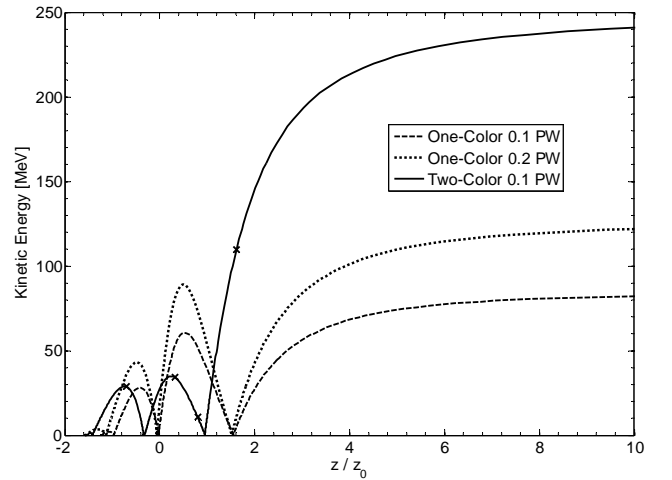

Fig. 3. Variation of kinetic energy with electron displacement of an electron hit by a pulse. In each case, $w_0 = 2$ μm and $\tau = 10$ fs, with all other parameters optimized. Crosses on the solid curve indicate the positions where the Fig. 4 plots are generated.

Applying the same method by which $\Delta E_{\lim}$ was formulated for the one-color beam gives us a theoretical gain limit for the two-color beam: $\Delta E_{\lim,TC} = 2^{1/2} \Delta E_{\lim}$ (given $P$). This may lead one to expect a two-color beam of total power $P$ and a one-color beam of power $2P$ to be comparable in electron acceleration capability. However, the former in fact significantly outperforms the latter for $P$ well above the electron acceleration threshold. As Fig. 3 shows, an electron in a 0.1 PW one-color beam slips through several accelerating and decelerating cycles, gaining and losing substantial amounts of energy, before finally entering its effectively final accelerating cycle. When the one-color beam is intensified to 0.2 PW (and optimum conditions re-computed), the final electron energy increases, but so have the heights of the intermediate energy peaks, which reduce net acceleration in this case by pushing back the position where the electron enters its final accelerating cycle. The two-color beam scheme achieves smaller intermediate peaks by varying the laser's interference pattern to increasingly favor acceleration as the electron moves forward past the focus (Fig. 4), adopting the acceleration-favoring Fig. 1(b) pattern only as the electron enters its effectively final accelerating cycle, instead of maintaining the same peak accelerating field at every position as the one-color beam does.

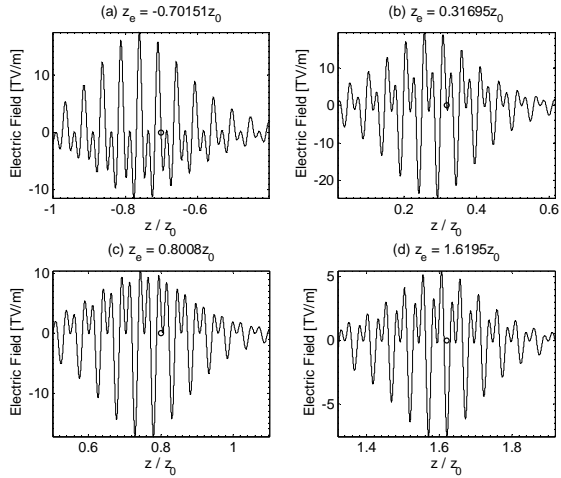

Fig. 4. $E_z$ profile of laser pulse at selected positions of the electron's trajectory for the two-color $P = 0.1$ PW case in Fig. 3. Circles at $z = z_e$ indicate the electron's position. (a), (b), (c) and (d) correspond respectively to the crosses in Fig. 3 from left to right.

Note that our scheme is fundamentally different from vacuum beat wave acceleration [10-13], which also uses a superposition of co-propagating laser beams, but which accelerates electrons by the beat wave arising from the $-e\vec{v} \times \vec{B}$ (ponderomotive force) term in the Lorentz force equation $\vec{F} = -e(\vec{E} + \vec{v} \times \vec{B})$, whereas our scheme accelerates electrons by the $-e\vec{E}$ term, using the Gouy phase shift to vary the overall interference pattern with position along the axis. Our scheme, which involves superposing pulsed laser beams an octave apart in central frequency and of equal Rayleigh range $z_0$ (which is necessary to leverage on the Gouy phase shift as we have done), also differs from a recent study [14] on electron acceleration which employs superposed radially-polarized laser beams that are neither an octave apart in frequency nor of equal $z_0$.

In conclusion, we have proposed and studied the direct acceleration of an electron in infinite vacuum by a two-color pulsed radially-polarized laser beam. This scheme exploits the presence of the Gouy phase shift to accelerate a stationary electron by over 90% of the one-color theoretical energy gain limit, more than twice of what is possible with a one-color beam of equal total energy and pulse duration. Future studies will examine the optimization of the two-color laser field for electron beam emittance and energy spread.

This work was financially supported by the National Science Foundation (NSF) grant NSF-018899-001 and the Agency for Science, Technology and Research (A*STAR), Singapore.

**References**
1. V. Malka, J. Faure, Y. A. Gauduel, E. Lefebvre, A. Rousse and K. T. Phuoc, Nat. Phys. **4**, 447-453 (2008).
2. T. Plettner, R. L. Byer, E. Colby, B. Cowan, C. M. Sears, J. E. Spencer and R. H. Siemann, Phys. Rev. Lett. **95**, 134801 (2005).
3. G. Malka, E. Lefebvre, and J. L. Miquel, Phys. Rev. Lett. **78**, 3314 (1997).
4. C. Varin, M. Piché, and M. A. Porras, Phys. Rev. E **71**, 026603 (2005).
5. Y. I. Salamin, Phys. Rev. A **73**, 043402 (2006).
6. Y. I. Salamin, Opt. Lett. **32**, 90-92 (2007).
7. A. Karmakar and A. Pukhov, Laser Part. Beams **25**, 371-377 (2007).
8. L. J. Wong and F. X. Kärtner, Opt. Express **18**, 25035–25051 (2010).
9. P.-L. Fortin, M. Piché, and C. Varin, J. Phys. B: At. Mol. Opt. Phys. **43** 025401 (2010).
10. H. Hora, Nature **333**, 337 (1988).
11. E. Esarey, P. Sprangle, and J. Krall, Phys. Rev. E **52**, 5443 (1995).
12. B. Hafizi, A. Ting, E. Esarey, P. Sprangle, and J. Krall, Phys. Rev. E **55**, 5924 (1997).
13. Y. I. Salamin, J. Phys. B: At. Mol. Opt. Phys. **38**, 4095 (2005).
14. Y. I. Salamin, Phys. Lett. A **375**, 795 (2011).

## Full References

1. V. Malka, J. Faure, Y. A. Gauduel, E. Lefebvre, A. Rousse and K. T. Phuoc, "Principles and applications of compact laser–plasma accelerators," Nat. Phys. **4**, 447-453 (2008).
2. T. Plettner, R. L. Byer, E. Colby, B. Cowan, C. M. S. Sears, J. E. Spencer and R. H. Siemann, "Visible-laser acceleration of relativistic electrons in a semi-finite vacuum," Phys. Rev. Lett. 95, 134801 (2005).
3. G. Malka, E. Lefebvre, and J. L. Miquel, "Experimental observation of electrons accelerated in vacuum to relativistic energies by a high-intensity laser," Phys. Rev. Lett. 78, 3314 (1997).
4. C. Varin, M. Piché, and M. A. Porras, "Acceleration of electrons from rest to GeV energies by ultrashort transverse magnetic laser pulses in free space," Phys. Rev. E **71**, 026603 (2005).
5. Y. I. Salamin, "Electron acceleration from rest in vacuum by an axicon Gaussian laser beam," Phys. Rev. A **73**, 043402 (2006).
6. Y. I. Salamin, "Mono-energetic GeV electrons from ionization in a radially-polarized laser beam ," Opt. Lett. **32**, 90-92 (2007).
7. A. Karmakar and A. Pukhov, "Collimated attosecond GeV electron bunches from ionization of high-Z material by radially polarized ultra-relativistic laser pulses," Laser Part. Beams **25**, 371-377 (2007).
8. L. J. Wong and F. X. Kärtner, "Direct acceleration of an electron in infinite vacuum by a pulsed radially-polarized laser beam," Opt. Express **18**, 25035–25051 (2010).
9. P.-L. Fortin, M. Piché, and C. Varin, "Direct-field electron acceleration with ultrafast radially-polarized laser beams: Scaling laws and optimization," J. Phys. B: At. Mol. Opt. Phys. **43** 025401 (2010).
10. H. Hora, "Particle acceleration by superposition of frequency-controlled laser pulses," Nature **333**, 337 (1988).
11. E. Esarey, P. Sprangle, and J. Krall, "Laser acceleration of electrons in vacuum," Phys. Rev. E **52**, 5443 (1995).

12. B. Hafizi, A. Ting, E. Esarey, P. Sprangle, and J. Krall, "Vacuum Beat Wave Acceleration," Phys. Rev. E **55**, 5924 (1997).

13. Y. I. Salamin, "Single-electron dynamics in a tightly focused laser beat wave: acceleration in vacuum," J. Phys. B: At. Mol. Opt. Phys. **38**, 4095 (2005).

14. Y. I. Salamin, "Direct acceleration by two interfering radially polarized laser beams," Phys. Lett. A **375**, 795 (2011).